\documentclass[twocolumn,superscriptaddress,showpacs,footinbib]{revtex4-1}  % for review and %submission
\PassOptionsToPackage{hyphens}{url}\usepackage{hyperref}
\usepackage{hyperref}
\usepackage{breakurl}
\usepackage{amssymb,amsmath,amsfonts}   % for math

\usepackage{bibentry}
\usepackage{natbib}
\usepackage{graphicx}  % needed for figures
\usepackage{dcolumn}   % needed for some tables
\usepackage{bm}        % for math
\usepackage{empheq}
\usepackage{enumerate}
\usepackage{slashed}
\usepackage{simplewick}
\usepackage{comment}
\usepackage{color}
\usepackage{soul}
\usepackage[english]{babel}
\usepackage{appendix}
\usepackage[utf8]{inputenc}

\usepackage{marginnote}
\usepackage[normalem]{ulem}

\newcommand{\ii}{{\rm{i}}}
\newcommand{\dd}{{\rm{d}}}

\newcommand{\nn}{\nonumber}

\newcommand{\eq}[1]{(\ref{#1})}
\renewcommand{\>}{\rangle}
\newcommand{\la}{\label}
\newcommand{\ba}{\begin{align}}
\newcommand{\ee}{\end{equation}}
\newcommand{\be}{\begin{equation}}

\def\12{\frac{1}{2}}
\def\3{\frac{1}{2\pi}}
\newcommand{\p}{\partial}
\newcommand{\en}{\end{align}}

\newcommand{\<}{\langle}

\usepackage{color}

\begin{document}

\setcounter{secnumdepth}{-1}

\title{Geometric adiabatic transport in quantum Hall states}
\author{S. Klevtsov}
\affiliation{Mathematisches Institut, Universit\"at zu K\"oln, Weyertal 86-90, 50931 K\"oln, Germany}
%\author{ T. Can}
%\affiliation{Simons Center for Geometry and Physics, Stony Brook University, Stony Brook, NY 11794, USA}  
\author{ P. Wiegmann}
 \affiliation{ Department of Physics, University of Chicago, 929 57th St, Chicago, IL 60637, USA}%Lines 

%  automatically or can be forced with \\
%  \email{Second.Author@institution.edu}
% \affiliation{%
%  \homepage{http://www.Second.institution.edu/~Charlie.Author}
% \affiliation{


\begin{abstract}
We argue that in addition to the Hall conductance and the nondissipative component of the viscous tensor, there exists a third independent transport coefficient, which is precisely quantized. It takes constant values along quantum Hall plateaus. We show that the new coefficient is the Chern number of a vector bundle over moduli space of surfaces of genus 2 or higher and therefore cannot change continuously along the plateau. As such, it does not transpire on a sphere or a torus. In the linear response theory, this coefficient determines intensive forces exerted on electronic fluid by adiabatic deformations of geometry and represents the effect of the gravitational anomaly. We also present the method of computing the transport coefficients for quantum Hall states.   \end{abstract}
\date{April 27, 2015}
\maketitle
%\nocitenames
\noindent{\it 1. Introduction}\; Quantum Hall (QH) states are distinguished by a precise quantization of the Hall conductance in materials with imprecisely known characteristics. A natural question is whether the Hall conductance is a unique quantized characteristic of the quantum Hall state. Are there any other independent transport coefficients precisely quantized on the QH plateaus?

Precise quantization in materials occurs when the transport is a nondissipative adiabatic process. The quantum Hall effect (QHE) is an example of a system where adiabatic conditions are in place. Namely, the low energy states are separated from the rest of the spectrum by a gap and adiabatic changes of parameters produce states with the flux. However, only adiabatic processes with the nontrivial first Chern class yield quantized transport coefficients.

In this Letter we show that apart from the Hall conductance there exist two more quantized transport coefficients, although at present only the former is experimentally accessible. One of these coefficients is the nondissipative component of viscous tensor introduced in Refs.\ \cite{Seiler1995,Levay,Levay1997}. Indications for the existence of another precise transport coefficient appeared recently in connection with the gravitational anomaly found in the context of QHE in Refs.\ \cite{Klevtsov2013,CLW,CLWBig,Klevtsov2014,clw2015,Abanov2013,GromovAbanovGalilei,framinganomaly,Read2015}, see also Ref.\ \cite{WZ2006}. 
 
Precise quantization on QH plateaus of the non-dissipative transport coefficients can be explained from two points of view. The first, topological explanation is through their relation to topological invariants, such as Chern numbers of vector bundles over the appropriate parameter space \cite{Thouless1982,*Thouless1985,Avron85,Avron1994}, see also \cite{Q}.  For example, in the case of the Hall conductance, the parameter space is spanned by  Aharonov-Bohm fluxes piercing through the handles of the Riemann surface. For the non-dissipative viscosity the relevant parameter space is the moduli space of complex structures on the torus \cite{Seiler1995,Levay,Levay1997}. In this Letter we show that the third coefficient shows up, when the parameter space is the moduli space of complex structures for the surfaces of genus 2 and higher.
 
For this reason, we discuss the precise transport for QH states on compact Riemann surfaces. We develop a general method to compute all three transport coefficients at once, with the emphasis on the third coefficient, which is the most subtle. We construct the topological invariants which are responsible for the precise quantization. Our method also sheds new light on the relation between the adiabatic transport in the QHE and conformal field theory. The method relies on holomorphic properties of the QH states.

The second view on the transport coefficients is via the local linear response theory. Although it does not establish quantization  \cite{Thouless1982,*Thouless1985,Avron85}, the linear response theory often provides a clearer physical interpretation. The third coefficient  we consider describes an intensive part of non-dissipative viscosity, which does not depend on the fluid density, and is an analog of Casimir forces. 
\smallskip

\noindent{\it 2. Electromotive adiabatic transport} \;   We begin with an example illustrating the quantization  of non-dissipative adiabatic transport and its relation to the linear response theory, which goes back to Refs.\ \cite{Laughlin1981,Thouless1982,*Thouless1985,Avron85,Avron1994}. We adopt the units in which  adiabatic parameters, transport coefficients  and adiabatic curvature are dimensionless. 

We consider the charge transport in QHE on a torus with Aharonov-Bohm (AB) fluxes  \(\varphi_a\) and \(\varphi_b\) along the $a$ and $b$ cycles. In the absence of dissipative diagonal components of the conductance matrix, the electromotive force (emf) \(\dot \varphi_b\) produces a current along the $a$ cycle  \(I_a=\frac 1{2\pi}\sigma_{ab}\,\dot\varphi_b\).  An adiabatic increase of the  AB-flux by the flux unit \(h/e\) transports the charge
   \(Q_a=\frac 1{2\pi}\int_0^{2\pi} \sigma_{ab} \,d\varphi_b\). The transported charge defines
the adiabatic transport coefficient \(\sigma_H=Q_a\) as an average of
the Hall conductance over the flux period.  A more general definition  \cite{Avron85} involves a non-dissipative conductance 2-form  \begin{align}\la{1}
\Omega=\3 \sigma_{ab} \,\delta\varphi_b\wedge
   \delta\varphi_a.
\end{align} 
Then the adiabatic transport coefficient is the average of this 2-form over a closed 2-cycle in the  parameter space (in this case, a torus \(T_\varphi:0\leq\varphi_a,\varphi_b<2\pi)\)
  \be\sigma_H=\3\int_{T_\varphi}\Omega.\la{11}\ee 
Following the arguments of Refs.\ \cite{Thouless1982,*Thouless1985,Avron85,TaoWu}, the conductance 2-form 
 \eq{1} is proportional to the adiabatic curvature 
  \begin{align}\la{2}  
\Omega=-\ii \,\<\delta\psi|\delta\psi\>.\end{align} 
In this formula \(\psi\) is a normalized ground state and \(\delta\psi\) its external derivative over the parameter space. In the fractional QH case, when the ground state on a closed surface is degenerate, the symbol $\<\delta\psi|\delta\psi\>$ includes the trace over all degenerate states divided by their total number. For example, on the torus there are $m=1/\nu$ Laughlin states \(\psi_1,\dots,\psi_m\), where $\nu$ is the filling fraction. Then \eq{2} reads \(  \Omega=-\ii \,\nu\sum_{r=1}^m\<\delta\psi_r|\delta\psi_r\>\). Mathematically the vector of the  ground states is a section of the rank $m$ hermitian vector bundle over the parameter space, whose first Chern number \(\frac {\ii}{2\pi }\,\sum_{r=1}^m\int \<\delta\psi_r|\delta\psi_r\>\) is an integer. Thus the conductance \eq{11} is quantized in units of the filling fraction $\nu$.

A   subtle difference between adiabatic transport and the conductance matrix was emphasized in  \cite{Thouless1982,Thouless1985,Avron1994}:  while the conductance 2-form \eq{1} may  fluctuate in mesoscopic systems, the adiabatic  transport coefficient \eq{11}  does not. The conductance 2-form consists of  a  precisely quantized part that saturates the adiabatic transport \eq{11}, and a
non-universal  exact 2-form which does not affect it. We  emphasize the difference by labeling the  precise adiabatic transport by a subscript \(H\),  like \(\sigma_H\)  in \eq{11} to  distinguish it from a non-precise linear response coefficient \(\sigma\) in \eq{1}.

This point reflects the difference
 of   approaches  of "effective action" \cite{Abanov2013,GromovAbanovGalilei,framinganomaly,Read2015} and the generating functional \cite{Klevtsov2013,CLW,CLWBig,Klevtsov2014}  with the adiabatic transport \eq{11}. The effective action is given by the integral of the adiabatic curvature \eq{2} over a surface in the parameter space  enclosed by the adiabatic process. Hence the entire conductance form \eq{1}, including the part, which is an exact 2-form, is relevant. In contrast, the adiabatic transport coefficient is given by the integration of the adiabatic curvature over a closed 2-cycle, as in \eq{11}. For this integral only the universal part of the conductance form \eq{11} is relevant, while the exact part of the 2-form does not contribute.

\smallskip 

\noindent{\it 3. Geometric adiabatic transport} \;In addition to the charge transport, there is another set of adiabatic parameters related to the deformations of geometry. In the seminal papers Avron,
Seiler, Zograf \cite{Seiler1995} and L\'evay \cite{Levay} computed the adiabatic transport associated with deformations of the complex modulus  \(\tau=\tau_1+\ii\tau_2\) of the torus. The modulus defines a complex structure via complex coordinate \(z=x+\tau y\). In these coordinates  the metric has the form \(ds^2= g_{z\bar z} |dz|^2\), with the diagonal components vanishing $g_{zz}=g_{\bar z\bar z}=0$, and \(g_{z\bar z}={V}/{\tau_2}\), where \(V\) is the area of the surface. 
  
An infinitesimal  change of the modulus \(\tau\to\tau+ \delta \tau\)  preserves the area but transforms the  metric
\begin{align}
\delta(ds^2)=\delta g_{z\bar z}|dz|^2+\delta g_{zz}(dz)^2+\delta g_{\bar z\bar z}(d\bar z)^2,\la{81}
\end{align}   
the differential $\delta\mu$ defined by the formula $ g_{z\bar z}^{-1}\delta g_{zz}=\delta\bar \mu,\quad g_{z\bar z}^{-1}\delta
g_{\bar z\bar z}=\delta\mu$ is called Beltrami
differential. If the variation preserves the volume of the surface, which we assume, then the variation of the conformal factor reads  $g_{z\bar z}^{-1}\delta g_{z\bar z}=2|\delta\mu|^2$. In the case of the torus \(\delta\mu=\frac{ \ii\delta\tau}{2\tau_2}\) does not depend on the coordinates. 

According to  \cite{Seiler1995,Levay}  the adiabatic
curvature is proportional to invariant area form on the moduli space \begin{align}
 \Omega=-2\ii\eta_H(\delta \mu\wedge  \delta\bar \mu), \quad \delta\mu=\frac{\ii \delta\tau}{2\tau_2}\la{4},
\end{align}  
where 
\(\eta_H\) is a universal 
transport coefficient.  The authors of \cite{Seiler1995} interpreted the 
 \((\hbar/V)\eta_H\) as a non-dissipative component of the viscosity. 

The computations of    \cite{Seiler1995,Levay} have been carried out for the integer QHE and on the torus.  They have been extended  in \cite{Tokatly2007,*Tokatly2009,Read2009,*Read2011} to the fractional QH-states, also  on the torus. It was shown that on the torus  the coefficient \(\eta_H\)  is extensive, i.\ e.\ proportional to the number of flux quanta 
 \be\la{99} \text{torus}:\quad \eta_H=\varsigma_H N_\Phi,\quad  N_{\Phi}=\3\int BdV.\ee
For the usual Laughlin states $\varsigma_H=1/4$. The parameter space $\mathcal M$  in the case of the torus is the fundamental domain of a certain subgroup of the modular  group, with the volume equal to \({\rm vol}\,\mathcal{M}=\ii \int \delta \mu\wedge \delta\bar \mu=\pi\). The integral of the adiabatic curvature \eq{4} over this space  \be\3\int_\mathcal{M}\Omega=-\frac {\eta_H}\pi {\rm vol}\,\mathcal{M}\la{R}\ee 
is the Chern number. Albeit non-integer it is a topological invariant which ensures precise quantization of $\eta_H$.

Now we turn to another  universal  coefficient.   We will show that the relation \eq{99}  acquires an intensive quantum correction  \eq{71}, which  becomes visible only on surfaces with genus two and higher.  Relation \eq{R} establishes its  preciseness.  We comment that the integer QHE  on compact surfaces with a constant negative curvature was first studied  in the important paper of
   L\'evay \cite{Levay1997}.  
\smallskip

\noindent
{\it\ 4. Geometric adiabatic transport - the main result}\;  We  state the main result first and then sketch its derivation. 

We briefly recall the basic notions of the moduli space of complex structures (Necessary
 facts about moduli space can be found in   \cite{BelavinJETP,*BelavinPL}).
We would like to consider deformations of the metric \eq{81}, which exclude unphysical coordinate reparameterizations, or diffeomorphisms, $z\to z+\epsilon(z,\bar z)$. These correspond to Beltrami differentials of the form $\p_{\bar z}\epsilon$, as follows from \eqref{81}. Physical deformations $\delta\bar\mu$ are orthogonal to diffeomorphisms with respect to the standard inner product: $\int_\Sigma (\p_{\bar z}\epsilon)\delta\bar\mu\, g_{z\bar z}dzd\bar z=0$. Thus they are given by holomorphic differentials
\be\p_{\bar z}(g_{z\bar
z}\delta\bar\mu)=0.\la{H}\ee 
For surfaces of a genus ${\rm g}\geq2$ there are $3{\rm g}-3$ independent holomorphic differentials \(\eta_l\). 
The corresponding Beltrami differential  \(\delta\mu=g_{z\bar z}^{-1}\sum_{l=1}^{3{\rm g}-3} \bar\eta_l\delta y_l\) is  characterized by   
\(3{\rm g}-3\) complex coordinates  \(\delta y_1,\dots,\delta y_{3{\rm g}-3}\) on the tangent space to the moduli space. On the torus the moduli space has complex dimension one. On the sphere the moduli space is just a point.

We recall the notion of the Weil-Petersson form on the moduli space. It is the form invariant with respect to a coordinate choice of the moduli space
\begin{align}\nn
\Omega_{WP}= \frac\ii{V}\int_\Sigma(\delta
 \mu\,\wedge\delta \bar \mu
\,)dV.
\end{align}
Here   \(dV=g_{z\bar z} dz d\bar z\) is the volume element of the surface $\Sigma$.
We will show that the {\it universal} part of the adiabatic curvature of QH-states on the moduli space is
\begin{align}
   \la{71}
\Omega=-2\eta_H\Omega_{WP}, \quad \eta_H=\varsigma_H  N_\Phi-\frac{c_H}{
24}\chi(\Sigma),
\end{align} 
where \(c_H\) is a new precise transport coefficient, and  \(\chi(\Sigma)=2\!-\!2{\rm
g}\) is the Euler characteristic of
the surface. 

We list the value of all  three precise coefficients for the spin-$j$ Laughlin states which we  defined in \cite{CLWBig,Klevtsov2014} 
\begin{align}\la{13}
\sigma_H=\nu,\; \varsigma_H=\frac 1{4}(1-2{j\nu}),\; c_H=1-\frac 3\nu(1-2j\nu)^2
\end{align}
and compute them below at once.   Notice that the value of $c_H$ for $\nu=1/3$ Laughlin state with $j=0$  is $c_H=-8$ and, remarkably $c_H=0$ at $j=1$ or $j=2$.  Also \(\varsigma_H\) may have any sign and  even  vanish for spin-$j$ states. In sec.\ 8 we identify the coefficient \(\varsigma_H\) and \(c_H\) with the background charge and the central charge of the relevant conformal field  theory.

The formula \eq{13}  generalizes the result of \cite{Seiler1995, Levay,Levay1997}, and also  \cite{Tokatly2007,*Tokatly2009,Read2009,*Read2011} to Laughlin states on an arbitrary surface. We emphasize that as an adiabatic  transport coefficient, \(c_H\) cannot be seen on the torus, since Eq.\eqref{71} then reduces to \eq{4} (cf.,\cite{Levay1997}).

In \cite{Read2009,*Read2011} it was argued 
that the extensive part   \(\varsigma_H N_\Phi\)  of $\eta_H$ in \eq{71} is linked
to the difference between the admissible number of  electrons and the magnetic
 flux. The relation between these two quantities has been suggested in   \cite{Wen1992a},
 \be N=\sigma_H N_\Phi+2\varsigma_H\chi(\Sigma).\la{15}\ee  
With the help of  \eq{15} we can write the non-dissipative viscosity  coefficient
in \eq{71} as
\begin{align}\la{123}
\eta_H=\frac {\varsigma_H}{\sigma_H}N-\frac{\chi(\Sigma)}{24}.
\end{align} 
We observe that the kinematic viscosity \(\hbar\eta_H/N\) receives a universal finite size
correction $-\frac{\chi(\Sigma)}{24}$. This  is analogous to the Casimir effect, where forces receive a volume independent contribution. The origin of this correction is the gravitational anomaly, as we demonstrate below.

The same arguments as in Sec.3  establish precise quantization of the coefficient $c_H$.  Since the integral of the  left hand side of Eq. \eq{71} over any closed 2-cycle in the moduli space is a topological invariant and the volume of these cycles in Weil-Petersson metric is  a rational number \cite{Wolpert1983}, the coefficients  \(\varsigma_H\) and  \(c_H\) are  precisely quantized. It is more difficult to establish the units in which these coefficients are quantized, since the fundamental domain is an orbifold.

We  emphasize that deformations of the metric which do not change the moduli, such as variations of the conformal factor \(g_{z\bar z}\) (Weyl transformations) or diffeomorphisms do not lead to new precise transport coefficients. 

\smallskip

 \noindent{\it 5. Defining relation for holomorphic states} \;The fundamental principle behind the 
precise quantization of the adiabatic transport coefficients is the holomorphic properties of states on the lowest Landau level. These are many-particle  states built from  one-particle  states   annihilated by the operator 
  $$ D^\dag=g_{z\bar
z}^{-1/2}(-\ii\p_{\bar z}-A_{\bar z}+j\omega_{\bar z}).$$
Here \(A_{\bar z}\) and \(\omega_{\bar z}\) are  complex  components of the (non-uniform)
gauge field and the spin connection, and the spin  \(j\) is a parameter. 
We recall that the spin connection is defined  such that its 
 exterior
 derivative is the (scalar) curvature. Similarly the exterior derivative of the gauge field  is the magnetic field.
$$\dd\omega=\frac 12RdV,\quad \dd A=B dV$$
  
  Thus the states are holomorphic functions of the coordinates if the gauge field and the spin connection are treated as adiabatic parameters. But there is more to it. Unnormalized states are also holomorphic functions on the space of adiabatic parameters, in our case the space of complex structure moduli. Under a deformation of  the metric
\eq{81} the operator \( D^\dag\)  deforms  holomorphically with \(\mu\)  as  \(  \delta D^\dag=\delta\mu
D    \) and
so do unnormalized wave-functions
\be\psi_r(z_1,\dots,z_N|\mu,\bar\mu)=
\frac 1{\sqrt{\mathcal{Z}[\mu,\bar\mu]}}{F_r(z_1,\dots,z_N|\mu)},\la{172}\ee
where the index $r$ labels degenerate fractional QH-states, for surfaces of genus $\rm g\geq1$. Laughlin states on the torus transform under a unitary representation of the appropriate subgroup of the modular group, see e.\ g.\ \cite{Hansson2014}. Hence the modular invariant normalization factor is the same for each state. We assume this property to hold on higher genus surfaces.  Under this assumption the common normalization  factor,  also known as a generating functional, determines  the  adiabatic curvature 
 \begin{align}
\Omega=\int_\Sigma  \bigl( \bar{\bf d}{\bf d}\log \mathcal{Z}\bigr)  dV,  \la{17}
\end{align}
 where  \({\bf d}=\delta\mu\frac{\delta}{\delta\mu}\) and \(\bar{\bf d}=\delta\bar
\mu\frac{\delta}{\delta\bar \mu}\) and similar for  AB fluxes. The formula \eq{17} follows directly from the definition \eq{2} and the property \eq{172}. 

The 
defining  relation \eq{17} is valid for any states with the holomorphic dependence
on  complex parameters. Such states occur  in a broad scope of physical systems,  notably
in conformal field theory, see e.g.  \cite{Bernard1988,*Bernard19882,Verlinde1989}.
\smallskip

\noindent{\it 6. Generating functional}\;  Thus, in order to compute the adiabatic curvature one needs to know the generating functional $\mathcal Z$.  For the Laughlin states it has been obtained
in Ref. \cite{CLWBig} (cf.\cite{CLW,Klevtsov2013,Klevtsov2014}). It consists
of two parts \be\log \mathcal{Z} =\log \mathcal{Z}_H+\mathcal{F}[B,R].\la{20}\ee The
first term is a bilinear combination of the gauge and spin connections $A_z$ and $\omega_z$. The second term is a local functional of the magnetic field, scalar curvature and their derivatives.
  
Assuming the transversal gauge \(\p_z A_{\bar z}+\p_{\bar z} A_z=\p_z \omega_{\bar z}+\p_{\bar z} \omega_z=0\), the result of \cite{CLWBig,CLW,Klevtsov2013,Klevtsov2014}  for the   first term in \eq{20} can  be written in the  matrix form
\begin{align}
 \la{19}&\log \mathcal{Z}_H=\frac{2}{\pi}\int (A_{\bar
 z} \ \omega_{\bar z}) \begin{pmatrix} \sigma_H
  & 2\varsigma_H \\ 2\varsigma_H & - \frac{c_H}{12} \end{pmatrix} \begin{pmatrix} A_{
 z} \\ \omega_{z} \end{pmatrix}dzd\bar z.
 \end{align}
 Now we have all the necessary data to compute the  adiabatic transport coefficients. They are  matrix elements of the Hessian of \(\log Z_H\) 
\begin{align}\nn
&\sigma_H=\frac \pi 2\frac{\delta^2\log\mathcal{Z}_H}{\delta A_z\delta A_{\bar z}},\; 2\varsigma_H =\frac \pi 2\frac{\delta^2\log\mathcal{Z}_H}{\delta\omega_z\delta A_{\bar z}},\\  
&-\frac{c_H}{12}=\frac \pi 2\frac{\delta^2\log\mathcal{Z}_H}{\delta\omega_z\delta \omega_{\bar z}}\nn
\end{align}

We will focus on the geometric transport.

Enforcing the condition \eq{H}, which excludes diffeomorphisms, the deformation of the spin connection  is composed of two distinct parts: the variation of the conformal factor $g_{z\bar z}$, which  deforms the  curvature but keeps the moduli fixed,  and the deformations along the moduli space. The variation of the generating functional \eq{20} with respect to the conformal factor is an exact one-form. Hence it does not contribute to the adiabatic transport. The only source for the adiabatic transport is the deformation of the moduli. Under these deformations  the spin connection  deforms as   \begin{align}
 {\bf
\bar d d}\omega_z= \frac 1 2\p_z (\delta \bar \mu\wedge \delta\mu).
\la{191}
\end{align}
 Then formulas (\ref{191}, \ref{19}) and  \eq{17} yield 
  \begin{align}\la{A}
\Omega=-\frac{\ii}{\pi}\int_\Sigma\left(\varsigma_HB-\frac{c_H}
 {48}  R\right)(\delta  \mu\,\wedge \delta\bar \mu\, )dV.
\end{align}

Restricting to constant magnetic field \(2\pi N_\Phi/V\) and the constant curvature \(4\pi\chi(\Sigma)/V\), the result \eq{71} immediately follows.

 The  second term in \eq{20}  contributes only to the exact part of the conductance 
2-form and therefore is not relevant for the precise adiabatic coefficients.  Nevertheless it can be computed for a model wave function \cite{CLWBig}. 
The leading \(1/N\) order of the functional \eq{20} 
  reads  \begin{align*} \mathcal{F}=&\frac 1{4\pi}
 \int\left(\frac {1}{12}-\frac{(1-2\nu)^2}{4\nu}\right)
 \left(\frac 12\Delta\log\mathcal{B}-R\right)\, dV
 +\\&\frac 1{4\pi}(1-2\nu)
 \int\mathcal{ B}\log\mathcal{B}\; dV,
 \end{align*}
 where  we denote  \(\mathcal{B}=B+\frac
  12(1-j) R\) \cite{CLWBig}.
\smallskip

\noindent
{\it 7. Linear response} \; Now we explain the physical meaning of transport coefficients in terms of the linear response theory.

  Consider,  an adiabatic  process
where the spin and the gauge connections  evolve along  an open path \(A(t),\,\omega(t
)\), while  magnetic field and the curvature remain unchanged. %where we used a short cut notation \(\sigma(z,z')=\frac \pi 2\frac{\delta^2\log\mathcal{Z}}{\delta A_z(z)\delta A_{\bar z}(z')} \). The %linear response functions include the universal contribution from $\mathcal{Z}_H$ in \eq{20}, related to the transport coefficients, and %also non-precise gradients corrections from \(\mathcal{F}\) in \eq{20}.
The adiabatic process  \(A(t),\,\omega(t
)\) gives rise to an electric field 
 \(E_z=\dot A_z\) and its gravitational counterpart \(\mathcal{E}_z=\frac 12\dot
\omega_z\). They in turn create an electric current and stress.  General relations connect the current \(I_i\) and the stress \(\pi_{ij}\) to the variation of the generating functional over the gauge potential and the metric whose components are treated as independent parameters and do not require the preservation of the  volume.  In complex coordinates  the relations for the current  \(I_idx^i=I_zdz+I_{\bar z} d\bar z\) and the stress (in units of $\hbar$) \(\pi_{ij}dx^idx^j=\pi_{zz}(dz)^2+\pi_{\bar z\bar z}(d\bar z)^2+2\pi_{z\bar z} dzd\bar z\) read
\begin{align}&
I_z=\frac 1{4\ii}\frac{d}{dt}\left(\frac{\delta\log\mathcal {Z}_H}{\delta
A_{\bar z}}\right),\la{190} \\
%\pi_{zz}\!=-\frac 1{8} \p_z\frac{d}{dt}\frac{\delta\log\mathcal {Z}_H}{\delta
%\omega_{\bar z}}
&\pi_{zz}\!=\!\frac 1{4\ii }\!\frac{d}{dt}\!\left(\!g_{z\bar z}\frac{\delta\log\mathcal {Z}_H}{\delta
g_{\bar z\bar z}}\!\right)=\!\frac 1{4\ii }\!\frac{d}{dt}\!\left(\frac{\delta\log\mathcal {Z}_H}{\delta
\mu}\!\right),\\
&\pi_{z\bar z}\!=\! \frac \ii2\!\frac{d}{dt}\!\left(\!g_{z\bar z}\frac{\delta\log\mathcal {Z}_H}{\delta
g_{z\bar z}}\!\right).\la{245}
\end{align}
Computation of the stress is simplified if these formulas are rewritten in terms of the variation over the spin connection.
Since 
\begin{align}\nn
\ii\delta\omega_{\bar z}&=g_{z\bar z}^{-1/2}\p_z(g_{z\bar z}^{1/2}\delta\mu)+ \p_z(g_{z\bar z}^{-1/2}\delta g_{z\bar z}^{1/2})
%\\
%&=\frac{\ii}{2} \delta \mu\,\p_z\log \sqrt g-\frac \ii 2\p_z\delta\log\sqrt  g 
\end{align} we obtain
 \begin{align}\nn
&\pi_{zz}=-\frac 1{8} \frac{d}{dt}\left(\p_z\frac{\delta\log\mathcal {Z}_H}{\delta
\omega_{\bar z}}\right),\\\nn
&\pi_{z\bar z}=-\ii \frac{d}{dt}{ \rm Im}\left(\p_{\bar z}\frac{\delta\log\mathcal {Z}_H}{\delta
\omega_{\bar z}}\right).
\end{align}
Then the formulas for the universal part of the current and the stress follow from  \eq{19} and \eq{190} \eq{245} 
\begin{align}
&I_z=\frac 1{2\pi\ii}\left(\nu E_z+2\varsigma_H\mathcal{E}_z\right),\la{241}\\
&\pi_{zz}=-\3\p_z\left(\varsigma_HE_z-\frac {c_H}{24}\mathcal{E}_z\right).
%=\nn\\&-\ii \frac {\varsigma_H}{\nu}\p_zI_z+\frac 1{12}\3\p_z\mathcal{E}_z
\la{242}
\end{align}
The trace of the stress \(\pi_{z\bar z}\) vanishes on divergence free fields.

  The formula \eq{241} extends the notion of the Hall conductance: the e.m. current
 is a sum 
 of Lorentz forces  caused by the electric and gravitational fields.  The formula \eq{242} introduces the remaining transport coefficient.

Equivalently, the transport coefficients could be seen from the relation between the density and the momentum on magnetic field and the curvature. These formulas are obtained  from the continuity equation and the momentum conservation law
\begin{align}\la{200}
\dot\rho+\nabla^i I_i=0,\quad
\dot P_i+\nabla^{j}\pi_{ji}=0.
\end{align}
 From \eq{245} and \eq{200}  we read the variational formulas for the  density  and the momentum 

\begin{align}
&\rho=\frac12{\rm Im}\left(\nabla_z\frac{\delta\log\mathcal {Z}_H}{\delta A_{ z}}\right),\quad \nabla_z=g^{z\bar z}\p_z,\la{2890}\\
&P_z=-\ii\p_{ z}\left(\frac{\delta\log\mathcal {Z}_H}{\delta
 g_{z\bar z}} \right).\la{288}
\end{align} 
Furthermore, integrating the density over a sub-volume of the fluid we obtain  the variational formula for the number of particles in the sub-volume.  Similarly, on the surface of revolution, the integral of the momentum with the Killing vector gives the variational formula for the angular momentum (in units of $\hbar$) of the fluid  sub-volume
\begin{align}
&N=\frac 1 2{\rm Im}\int \left(\nabla_z\frac{\delta\log\mathcal {Z}_H}{\delta A_{ z}}\right)dV,\\
&\mathrm{L}=-\int \left(\frac{\delta\log\mathcal {Z}_H}{\delta
 g_{z\bar z}} \right)dV.
\end{align}
Then using (\ref{241}-\ref{200})   we obtain the extensions of the  St\v reda formula expressing the density and the momentum in terms of magnetic field and the curvature
\begin{align}
& \rho=\frac 1{2\pi}\left(\nu B+{\varsigma_H} R\right),\la{30}\\
&P_z=\frac {1}{\ii\pi} \p_{z}\left(\varsigma_H\, B -   \frac{c_H}{48}
 R\right),
\la{222}
\end{align} 
 and the formula expressing the number of particles  and the angular momentum  of a sub-volume of the fluid through the mean magnetic field  and the mean density \ \begin{align}
& N=\frac 1{2\pi }\int\left(\nu  B+{\varsigma_H}  R\right)dV,\la{32}\\
& \mathrm{L} =  -\frac {1}{\pi }\int\left(\varsigma_H   B- \frac{c_H}{48}
{ R}\right)dV.\la{33}
\end{align} 
The pair of formulas (\ref{241},\ref{242}), (\ref{30},\ref{222}) and (\ref{32},\ref{33}) further illustrate the meaning of the geometric  transport coefficients.

%Eqs. \eq{222} and \eq{71}  check the general relation 
%\begin{align}\mathrm{L}=-2\eta_H\nn
%\end{align}
%between the angular momentum of the fluid and the transport coefficient \(\eta_H\)  \cite{Read2009,*Read2011}.
 \smallskip

 \noindent{\it 8. QH-state as a string of  vertex  operators}\;  As we have seen, the generating functional (\ref{19}) is the central object of the theory of transport in QH-states. In the remaining part of the paper we outline one of available methods  to obtain it.  The method is based on the  construction of Ref.  \cite{Moore1991}  where   QH-states are expressed  by  a  string  of \(N\)  vertex operators in a relevant  field theory, coupled to the magnetic field. This approach  has been recently developed in  \cite{Klevtsov2014} (see also \cite{CLWBig}). We illustrate  this method  for the Laughlin spin \(j\)-states, defined in \cite{CLWBig,Klevtsov2014}, and assume no AB-fluxes. 
  
We look for a field theory which represents  the unnormalized part of the Laughlin wave function. 
Since this state consists of only one type of particles, it is described by  one  Gaussian field 
 \(\Phi\) coupled to the magnetic field and curvature \footnote{This action appeared in QH context in Ref.\ \cite{Kvorning}. We thank T.~H.~Hansson for pointing this out to us.},
\begin{align}
S[\Phi]\!=\!\!\frac {\sigma_H}{8\pi }\!\!\int(\nabla\Phi)^2 dV+\!\frac {\ii}{2\pi
}\!\!\int \!(\sigma_H B\!+\varsigma_H R)\,\Phi\,  dV\la{21},
\end{align} 
where the coupling constants  \(\sigma_H,\varsigma_H\) are fixed   by the   
requirements:

 (i) An electron is represented by a holomorphic
primary operator \(V(z)\) with electric charge 1. Identification of the 
vertex operator with $e^{\ii\Phi(z,\bar z)}=V(z)V(\bar  z)$ fixes  the coupling to the gauge field. 

(ii) The  OPE of two operators should satisfy  \(V(z_1)V(z_2)\sim(z_1-z_2)^{m}\)  as $z_1\to z_2$, where $m=1/\nu$.
This condition  determines \(\sigma_H=\nu\) in \eq{21}.

(iii) In the spin-$j$ Laughlin state, a particle has the conformal spin \(j\) \cite{CLWBig,Klevtsov2014}. This  state is a generalization of the usual  Laughlin state, for which $j=0$.  Since the state is holomorphic,
its conformal dimension also equals to \(j\).  We recall that the conformal
 dimension of the vertex operator \(e^{\ii a\Phi}\) with respect
to the action \eq{21} is  $\frac
a{2\sigma_H}(a-4\varsigma_H).$ Choosing \(a=1\) and the dimension to equal to its spin $j$, we obtain    \(\varsigma_H=\frac 1{4}(1-2{j\nu})\) as in \eq{13}.
This condition fixes the parameters of the spin-$j$ Laughlin state. 
The central charge of such theory 
 $$c_H=1-48\frac{\varsigma_H^2}{\sigma_H}$$
 is given by \eq{13}.
 
Now let us compute the unnormalized correlation function of  a string of  vertex operators $e^{\ii\Phi}$, following  \cite{Klevtsov2014}. We reproduce the unnormalized   Laughlin wave-function  \eq{172} 
 \begin{align} \frac 1{\mathcal{Z}_G}\!\int\! \left[\prod_{i=1}^Ne^{\ii\Phi(z_i,\bar z_i)}\right]e^{\!-\!S[\Phi]}D\Phi=
|F(z_1,\dots,z_N)|^2 \la{224}
\end{align}
 For example, on the sphere the state reads 
\begin{align}\la{23}F(z_1,\dots,z_N)\!=\prod_{i<j}^N(z_i\!-\!z_j)^{m}
e^{\!\frac 12\sum_{i=1}^N Q(z_i,\bar z_i)},
\end{align}
where   the  potential \(Q\) is such that  \(\p_{\bar z} Q= 2\ii(A_{\bar z}-j\omega_{\bar z})\).

The factor   \(\mathcal{Z}_G\)  in \eq{224} is 
\begin{align}
\mathcal{Z}_G= {[{\rm Det}(-\Delta)]^{-\frac 12}}
e^{-\frac{2}{\pi\sigma_H}\int|\left (\sigma_H A_{ z}+ 2\varsigma_H\omega_{ z}\right)|^2dzd\bar z},\la{D}
\end{align}
where ${\rm Det}(-\Delta)$ is the spectral determinant of the Laplace  operator.
The next step is to  integrate over positions of particles and use the relation $\int |F|^2 dV_1\dots dV_N=\mathcal{Z}$, where $\mathcal {Z}$ is the normalization factor in \eq{172}. We denote  $e^{\mathcal{F}}=\int \left[\int e^{\ii\Phi(z,\bar z)}dV\right]^N e^{-S[\Phi]}D\Phi$. Then the integration over the coordinates yields
\be 
\label{291}
e^{\mathcal{F}}=\mathcal {Z}\cdot\mathcal {Z}_G.
\ee
 To complete the argument we notice  that the  l.h.s. of \eq{291}
depends locally on the curvature  and does not depend on moduli.
Comparing to (\ref{20}) we obtain the main relation    \begin{align}
\mathcal{Z}_H^{-1}=\mathcal{Z}_G.\la{x}
\end{align}  
It  remains  to recall the value of the spectral determinant  of the Laplace operator in \eq{D}. Up to metric independent terms it is given by the formula of Polyakov \cite{Polyakov1981}
\begin{align}\log {\rm Det}(-\Delta)=-\frac 1{3\pi}\int |\omega_z|^2 dzd\bar z.
\end{align} 
It represents the effect of the gravitational anomaly. This term corresponds to 1 in the formula for \(c_H\)  \eq{13} and is responsible for the finite size correction to the non-dissipative viscosity  \eq{123}.
The result for the generating functional of QH states \eq{19} follows from \eq{x}.
\smallskip

We acknowledge that this work has been initiated by discussion with T.~Can  whom we are grateful for insights, discussions and help. 
We would like to thank A.~G.~Abanov,  D.~Bernard, A.~Cappelli, Y.~H.~Chui, M.~Douglas, A.~Gromov,  M.~Laskin, N.~Read,  D.~T.~Son and S.~Zelditch for useful discussions. We thank A.~G.~Abanov, A.~Gromov, B.~Hanin, and N.~Read for their comments on the manuscript. The work of S.K. was supported in part by the Max Delbr\"uck prize for junior researchers at the University of Cologne, the Humboldt fellowship for postdoctoral researchers, Grants No. NSh-1500.2014.2 and No. RFBR 14-01-00547. The work of P. W. was supported in part by NSF Grants No. DMS-1206648, No. DMS-1156656, and No. DMR-MRSEC-1420709 and PVE grant from the CNPq-Brazil Science Without Boarders Program. We would like to acknowledge support from the Simons Center for Geometry and Physics, Stony Brook University, where this paper was finalized.

{\it Note added.}--Recently, we became aware of a recent paper \cite{Read2015} on the third coefficient $c$, where the holomorphic properties were employed to compute the geometric part of the adiabatic curvature.

\bibliography{adiabaticphase_refs}
\end{document}